\documentclass[11pt,a4paper]{article}

\usepackage[utf8]{inputenc}
\usepackage[margin=1in]{geometry}
\usepackage{amsmath,amssymb,amsfonts}
\usepackage{graphicx}
\usepackage{hyperref}
\usepackage{cite}
\usepackage{booktabs}
\usepackage[version=4]{mhchem}
\usepackage{siunitx}
\DeclareSIUnit\Molar{M}

\hypersetup{
    colorlinks=true,
    linkcolor=blue,
    citecolor=blue,
    urlcolor=blue
}

\title{\textbf{Self-organization in Chiral Gliding Filaments}}

\title{\textbf{Self-organization in Chiral Gliding Filaments}}

\author{%
  Josue Yaedalm Son$^{1,2,3}$, 
  Hyejeong Kim$^{1,2,3}$, and 
  Vahid Nasirimarekani$^{2,3,*}$\\[2ex]
  \parbox{\linewidth}{\centering \small
    $^{1}$School of Mechanical Engineering, Korea University, Seoul, 02841, South Korea\\[0.5ex]
    $^{2}$Max Planck Institute for Dynamics and Self-Organization, Am Fassberg 17, 37077, Göttingen, Germany\\[0.5ex]
    $^{3}$Laboratory of Fluid Physics, Pattern Formation and Biocomplexity, Am Fassberg 17, 37077, Göttingen, Germany
  }\\[1.5ex]
  \small $^{*}$Corresponding author: \texttt{vahid.nasirimarekani@ds.mpg.de}
}

\date{}

\begin{document}

\maketitle

\begin{abstract}
Filamentous cyanobacteria form extensive kilometer-scale mats in aquatic environments, yet the physical mechanisms driving their macroscopic self-organization remain elusive. Here, we investigate the dynamics of filamentous cyanobacterial assemblies within a quasi-two-dimensional confinement. Our observations reveal that chiral gliding drives active tangling between individual filaments, inducing rapid network contraction and the formation of densely packed aggregates. By quantifying single-contact forces, we demonstrate that these mechanical interactions generate significant contractile stresses. Together, our findings suggest that chiral gliding has evolved as a physical mechanism to maximize inter-filament interactions and drive large-scale collective organization.
\end{abstract}

\section*{Introduction}

Organisms or microorganism in nature can self-organize themselves into larger, macroscopic assemblies as seen in active matter such as bird flocks, fish schooling or bacteria colonies~\cite{marchetti2013hydrodynamics,vicsek2012collective,ramaswamy2010mechanics,shankar2022topological}. Filamentous network of active matter exhibit intricate interactions that lead to emergent pattern formation~\cite{vliegenthart2020filamentous,tsang2017dynamic,nasirimarekani2021tuning,dunajova2023chiral,nasirimarekani2022active}, chemical signaling~\cite{proust2011strigolactones,forgacs1995possible}, and environmental adaptation~\cite{karasz2022conditional,banerjee2020actin}. Mechanical forces play a key role in filament buckling and tension-driven reorientation which then would contribute to spatial organization~\cite{sanchez2012spontaneous,vliegenthart2020filamentous,deblais2023worm}. Such systems manifest how simple filamentous units, through local interactions, generate complex structures with adaptive functions~\cite{de1971reptation,weiner2020mechanics}.

One prominent example of self-organizing organisms is cyanobacteria, a type of photosynthetic bacteria that can form kilometers large, visible structures like mats in aquatic environments~\cite{stal2012cyanobacterial,whitton2007ecology, sciuto2015cyanobacteria}. The cyanobacteria population mainly form of single cells or filamentous cyanobacteria~\cite{schirrmeister2011origin,demoulin2019cyanobacteria,vidal2021introduction}. The filaments usually have a diameter in micrometers and are up to a few millimeters long~\cite{vidal2021introduction}. The size difference of 6-7 orders of magnitude between the filaments and the mats raises the following questions: (i) how can these microscopic filaments form such large assemblies, and (ii) what are the mechanisms underlying the filament-filament interaction and the stability of the final assemblies?

Beyond their fundamental biological role and ecological impact in diverse environments~\cite{havens2008cyanobacteria,paerl1996microscale,rastogi2015bloom}, cyanobacterial filaments are particularly intriguing due to their ability to function far from thermodynamic equilibrium~\cite{abbaspour2023effects, faluweki2023active, faluweki2022structural}. These active filaments harness internal or external energy sources, driving self-organization and giving rise to complex collective behaviors~\cite{faluweki2023active}. As a result, studying such systems provides a unique interdisciplinary bridge between physics and biology, offering valuable insights into the intricate mechanisms underlying microbial pattern formation. From a physics perspective, a crucial first step in unraveling these dynamics is to gain a deeper understanding of the interactions between individual filaments.

A sub-specie of filamentous cyanobacteria secretes extracellular polymeric substances (EPS), which act as a slime, binding cells together and anchoring them to surfaces such as rocks, sediments, and submerged vegetation~\cite{hoiczyk1995envelope,bertocchi1990polysaccharides}. These filaments glide easily on a hydrated substrate and form large clusters in an aqueous volume. An example of these filaments is the \textit{Lyngbya} family, which consists of flexible and elastic filaments~\cite{faluweki2022structural}. We previously studied a sub-species, \textit{Lyngbya lagerheimii}, and observed rotation of the filament along its long axis, which results in chiral gliding~\cite{vilfan2026chiral}. This rotational motion causes the filaments to bend in a right-handed manner on a substrate. In this article, we explore how these rotational and chiral dynamics govern filament-filament interactions within dense populations, driving emergent macroscopic self-organization.

\begin{figure}[htbp]
    \centering
    \includegraphics[width=\textwidth]{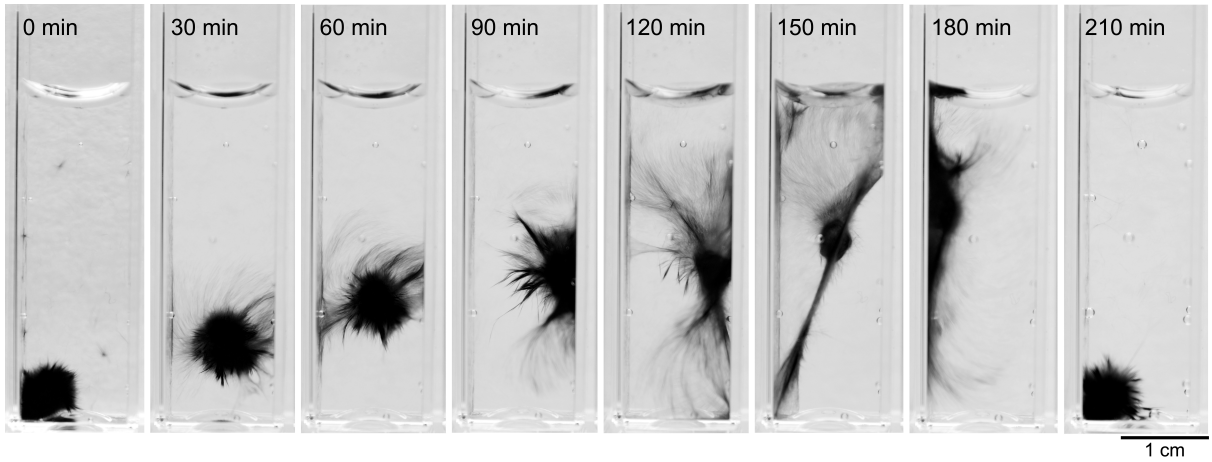}
    \caption{Dynamic assemblies of gliding filamentous cyanobacteria, (\textit{Lyngbya lagerheimii}). 
    Time-lapse images of the bacteria network showing the dynamic nature of the network, as partial disassembly and re-assembly of the filaments.}
    \label{fig:fig1}
\end{figure}

The motivation for this work stems from a direct observation: when suspended in water or a three-dimensional volume, the filaments self-assemble into tightly packed structures (Figure~\ref{fig:fig1}). This arrangement is highly dynamic, exhibiting floating and sinking cycles while maintaining its interconnected network architecture. Remarkably, the filaments recover their compact assembly following partial disassembly (as observed in the 120- and 150-minute snapshots). Although the dynamic nature of these assemblies originates from their biological cycles in aquatic environments~\cite{oliver1994floating,hense2006towards}, we contend that physical principles ultimately govern the inter-filament interactions and the resulting macroscopic patterns.

We hypothesize that cluster formation from individual filaments is driven by strong inter-filament interactions. An interaction involving the exertion of local or global forces between them. It was therefore thought that microscopic observation of filament-filament interactions would offer critical insights into the fundamental physical principles governing their self-organization. In this regard, here, we investigate the dynamic assembly and self-organization of filamentous cyanobacteria (\textit{Lyngbya lagerheimii}) within a quasi-two-dimensional confinement.
Therefore, we aim to observe and quantify the aggregation behavior in the filamentous network and identify the mechanism influencing the formation of compact, densely packed assemblies.

\section*{Results and Discussion}

\subsection*{Active Contraction and Expansion of the Filamentous Network}

\begin{figure}[htbp]
    \centering
    \includegraphics[width=0.9\textwidth]{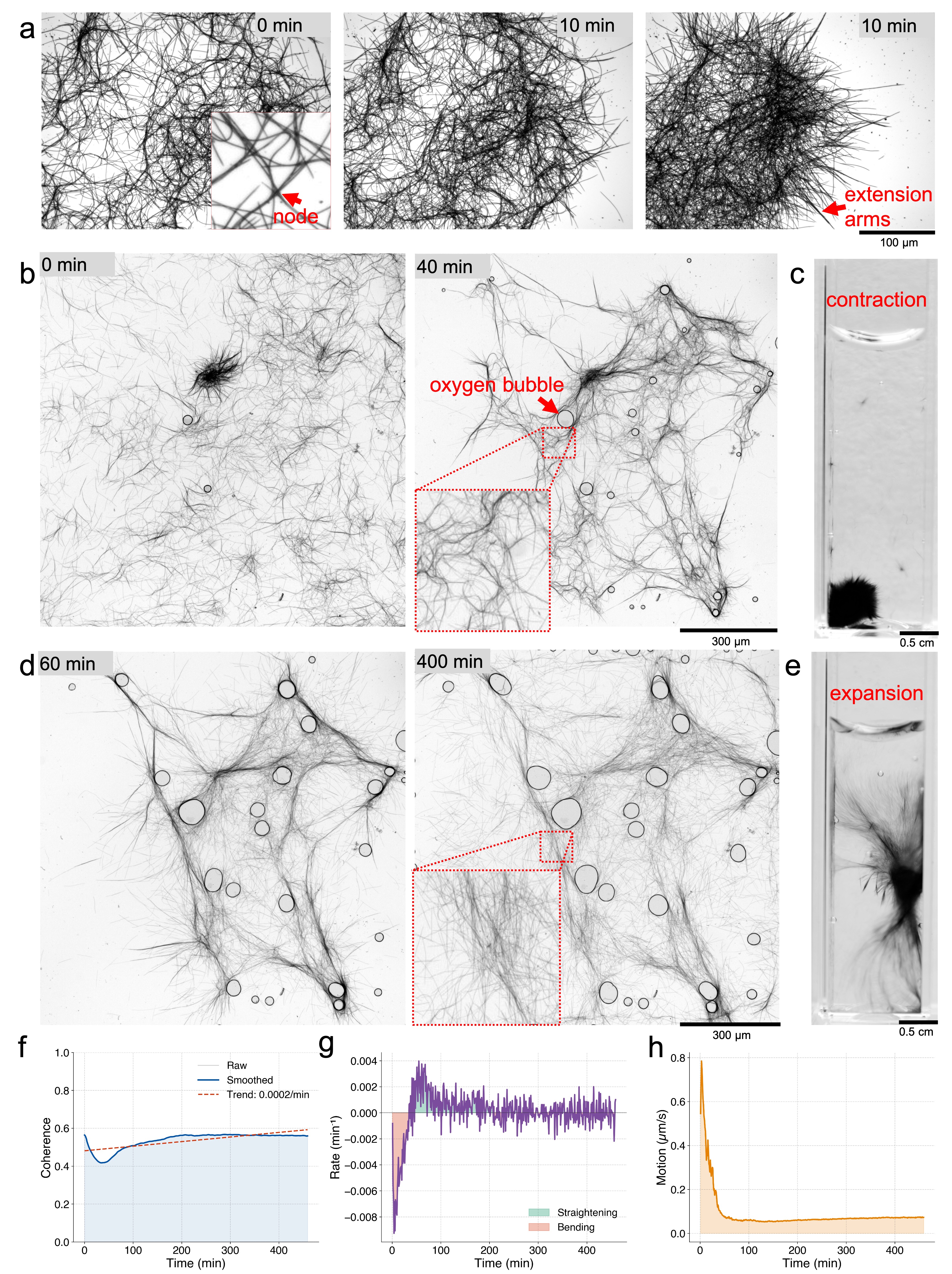}
    \caption{Contraction dynamics of the filamentous cyanobacteria network in quasi-two-dimensional confinement. a) Time-lapse images showing contraction of an interconnected filamentous network over 10 minutes. b,c) Contraction cycle of the network in quasi-2D confinement and in the cuvette experiment. Inset shows curved, flexible filaments during contraction. d,e) Expansion cycle of the network in quasi-2D confinement and in the cuvette experiment. Inset shows straightened filaments during expansion. f) Filament alignment coherence over time. Increasing coherence indicates filament straightening, while decreasing coherence corresponds to bending activity. g) Straightening rate (temporal derivative of coherence). Positive peaks indicate active expansion/straightening events; negative troughs indicate contraction/bending events. Shaded regions denote expansion (green) and contraction (orange) phases. h) Filament motion activity measured from optical flow analysis, with peak activity observed in the contraction phase.}
    \label{fig2}
\end{figure}

We subjected filamentous suspensions to quasi-two-dimensional confinement (Appendix Figure 1). Initially, individual filaments were homogeneously distributed in the bulk but rapidly interacted to form a highly cross-linked network (Figure~\ref{fig2}a, Video S1). These filaments coalesced into thicker bundles, which subsequently established an interconnected architecture featuring clearly visible nodes. Over time, the network underwent pronounced contraction, producing denser bundles as local filament concentration increased. Eventually, the entire structure compacted, with long bundles extending outward from the dense core (Figure~\ref{fig2}a, 20~min).

Prolonged observation of the entire network within a larger chamber revealed a biphasic behavior: the network first contracts and subsequently expands as filaments migrate outward from the aggregate (Figure~\ref{fig2}b-d, Video S2). Furthermore, the photosynthetic activity of the filaments drives the nucleation and growth of oxygen bubbles within the confinement, which act as physical obstacles that impede network contraction. In three-dimensional volumes, we infer that oxygen bubbles accumulate at the center of the cluster (Figure~\ref{fig:fig1}, Appendix Figure 2), giving rise to cyclic floating and sinking (bouncing) dynamics similar to those observed in other photosynthetic bacterial species~\cite{sanchez2020origin,soo2017origins}.

The morphology of individual filaments and bundles shifts significantly across contraction and expansion cycles. During contraction, the filaments appear flexible and adopt curved configurations (Figure~\ref{fig2}b, inset, Video S3), whereas during expansion, they appear more rigid and nearly straight (Figure~\ref{fig2}c, inset). This suggests that the filaments might modulate their effective bending stiffness during the cycles, though measuring stiffness directly in this setup remains challenging. To quantify these morphological changes, we performed a texture-based analysis of filament alignment and motion dynamics (Figure~\ref{fig2}f-h). The filament alignment coherence, a measure of structural order and straightness, exhibited clear temporal modulation corresponding to each phase (Figure~\ref{fig2}f). Coherence decreased during contraction as the network compacted and filaments bent, and increased during expansion as they dispersed and straightened. Furthermore, the straightening rate, derived from the temporal derivative of coherence, revealed distinct negative troughs during contraction (reflecting active bending) and positive peaks during expansion (reflecting filament straightening) (Figure~\ref{fig2}g).

Concurrent measurements of filament motion activity, quantified via optical flow analysis, revealed elevated motility during both contraction and expansion phases, exhibiting a mean speed of $0.1 \pm 0.01$ $\mu\text{m}\cdot\text{s}^{-1}$ and a maximum value of $0.78$ $\mu\text{m}\cdot\text{s}^{-1}$ (Figure~\ref{fig2}h). Motion activity peaked specifically during the contraction period, pointing to a robust active aggregation mechanism. This finding directly addresses the central question of our study: what underlying mechanism governs the formation of such compact and stable assemblies?

We assume that the observed contraction and expansion cycles, occurring in both 3D and 2D, are fundamentally mediated by strong filament-filament interactions, which drive both initial compaction and subsequent dispersion. It is worth noting that directional reversal~\cite{abbaspour2023effects,hoiczyk2000gliding,shepard2010undirected,vilfan2026chiral} is an intrinsic characteristic of these gliding filaments, allowing them to dynamically alter their trajectories. Moreover, the structural straightening observed during expansion likely facilitates faster spreading over a larger surface area. These dynamics motivate a closer examination of filament-filament interactions at the single-pair level to elucidate the physical principles governing self-organization in this system.

\subsection*{Chiral Gliding Interactions Drive Helical Tangling}

\begin{figure}[htbp]
    \centering
    \includegraphics[width=\textwidth]{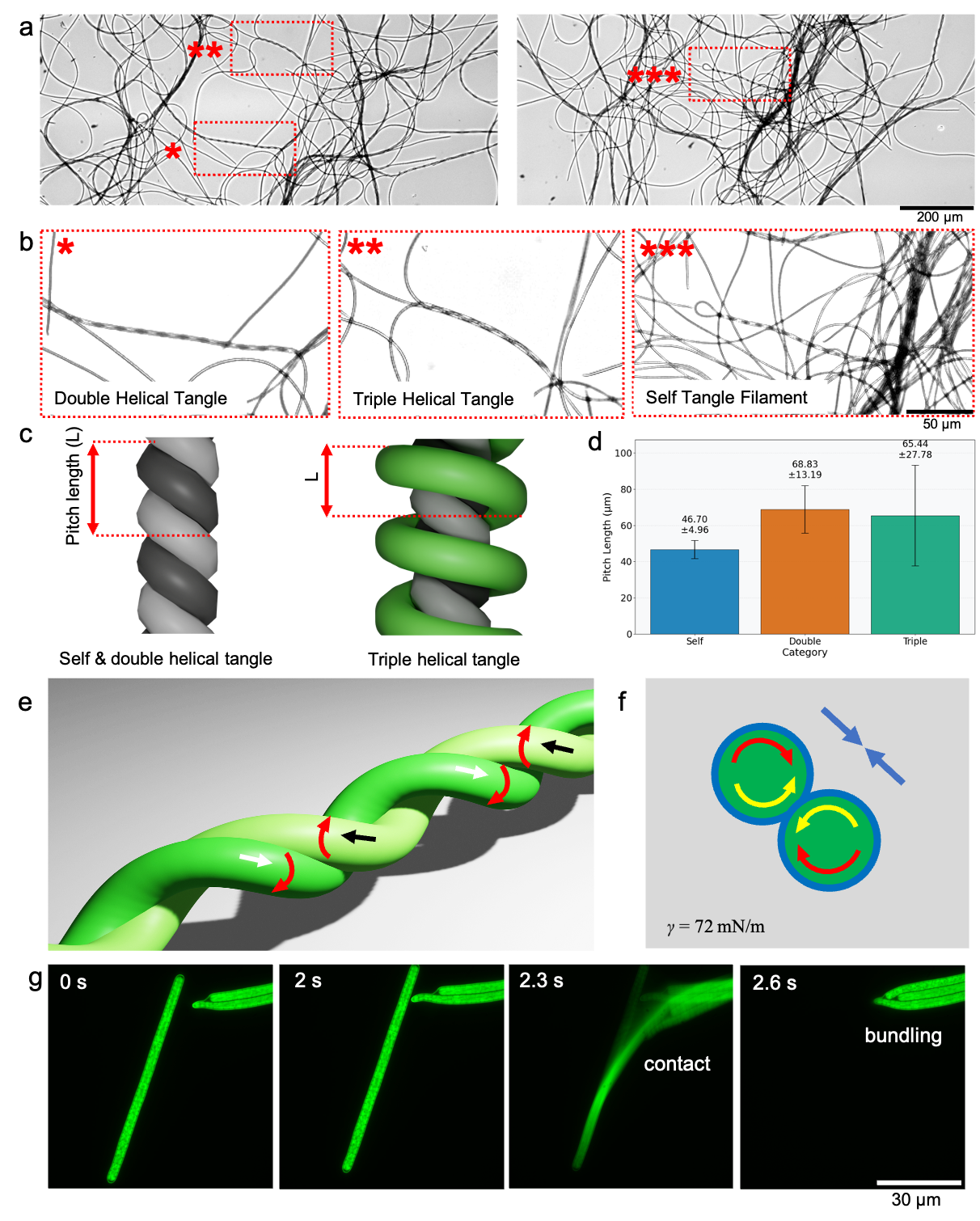}
    \caption{Formation of helical tangles in chiral gliding cyanobacterial filaments (\textit{Lyngbya lagerheimii}). a) Filamentous network showing rapid contraction with prominent helical tangles (highlighted regions). b) Zoomed views of helical tangles formed by two and three filaments, alongside a self-tangling case. c,d) Schematic representation of pitch length variations across different tangling configurations, with measured values ($n = 10$ for each case). e) Schematic illustration of two interacting filaments, where each acts as a chiral substrate inducing mutual bending. f) Schematic illustrating how the contact of slime tubes surrounding adjacent filaments eliminates slime–water interfacial area, lowering total free energy. g) Experimental observation of strong attractive interactions leading to filament bundling.}
    \label{fig3}
\end{figure}

Microscopic examination of gliding filaments during the early stages of network contraction reveals the intricate dynamics of helical tangle formation (Figure~\ref{fig3}a). Closer inspection highlights distinct structural motifs, including double helical tangles where two filaments intertwine in a well-defined configuration, and single filaments forming self-tangles (coils) characterized by knot-like structures (Figure~\ref{fig3}b,d, Videos S4–S5). Additionally, bundles comprising three or more filaments engage in complex entanglements, where a tangled bundle can serve as a gliding substrate for additional filaments (Figure~\ref{fig3}c, Video S6).

Quantitative analysis reveals distinct differences in the pitch lengths of these tangled regions. Self-tangled filaments exhibit comparatively shorter pitch lengths, averaging $46.70 \pm 4.96$ $\mu\text{m}$, whereas double helical tangles display longer pitch lengths, averaging $68.83 \pm 13.19$ $\mu\text{m}$ (Figure~\ref{fig3}d). This variation points to distinct mechanical constraints governing single-filament coiling versus two-filament intertwining.

Helical tangles indicate active, reciprocal interaction between filaments rather than passive sliding (Appendix Figure 3). We hypothesize that in a double tangle, each filament acts as a physical substrate for the other, causing one filament to bend helically around its partner (Figure~\ref{fig3}e). Furthermore, when the slime tubes secreted by adjacent filaments come into direct contact, part of the slime–water interfacial area is eliminated, thereby reducing the total interfacial free energy of the system. This energetically favorable contact produces capillary adhesion that holds the filaments together and promotes stable bundle formation. Such bundle formation is similarly observed under two-dimensional confinement, where geometric constraints prevent three-dimensional tangling (Figure~\ref{fig3}f,g, Video S7).

The dynamic nature of filament tangling plays a critical role in driving rapid network contraction and forming compact assemblies, a phenomenon reminiscent of collective entanglements observed in other active systems such as worm blobs~\cite{patil2023ultrafast,backholm2014tangling,deblais2023worm,ozkan2021collective}. These tangles align filaments into robust bundles that act as active contractile elements (Appendix Figure 4), pulling the interconnected network inward. More broadly, structural entanglement profoundly influences material properties across both biological and synthetic systems. In biological tissues, the intertwining of filamentous proteins such as collagen~\cite{ramachandran1955structure,shoulders2009collagen} and elastin~\cite{mammi1968evidence} balances mechanical strength and flexibility, supporting functions like tissue elasticity. Similarly, entangled polymer networks exhibit viscoelastic behavior, providing essential resilience and energy dissipation~\cite{keller2003slow}. Analogously, we conclude that these intertwined structures are fundamental in determining the mechanical, dynamic, and functional properties of the cyanobacterial network.

\subsection*{Filament-filament interactions generate contractile forces}

\begin{figure}[htbp]
    \centering
    \includegraphics[width=\textwidth]{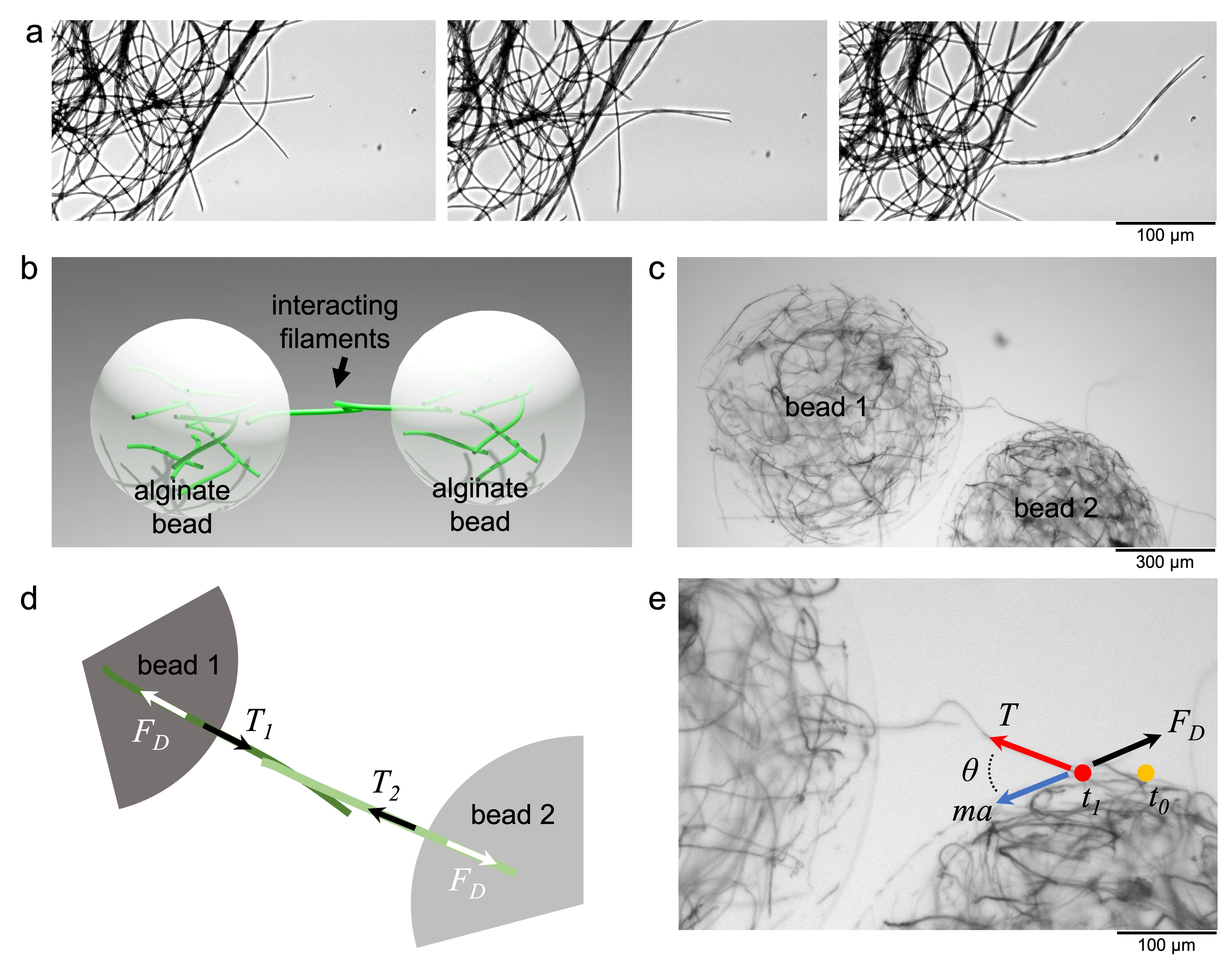}
    \caption{Filament-filament interactions generate contractile forces. a) Extension arms at the free edges of the cyanobacterial network. b,c) Experimental design showing cyanobacteria-encapsulated alginate beads (CABs) in BG-11 medium, where protruding filament arms engage, bind, and generate force to induce bead rotation and aggregation. d) Force balance diagram for translational motion, showing filament tension $T$, drag force $F_D$, and mass acceleration $ma$. e) Representation of force-induced bead rotation and torque balance.}
    \label{fig5}
\end{figure}

Tangling of filaments was also observed at the free edges of the cyanobacterial network, which are characterized by prominent extension arms (Figure~\ref{fig5}a). These extension arms provide an ideal opportunity to study interactions between individual filaments and their surrounding environment. 

To isolate and quantify these interactions, we developed an experimental approach by encapsulating cyanobacterial filaments within alginate hydrogel beads. This encapsulation allowed extension arms to protrude beyond the bead boundary, enabling us to examine how individual filament arms interact with neighboring structures (Figure~\ref{fig5}b). Our observations revealed that extended filaments from different beads interact and bind, resulting in stable inter-bead attachment (Figure~\ref{fig5}b,c, Video S8). Moreover, these interactions generated sufficient force to induce bead rotation, pulling the beads together and causing observable bead deformation (Appendix Figure 5, Video S9).

We quantified the force exerted by a single filament arm by tracking the motion of cyanobacteria-encapsulated alginate beads from time-lapse imaging (Figure~\ref{fig5}d,e). Depending on the interaction geometry, the filament induced either translational or rotational motion of the bead (Appendix Figure 6). For translational motion, the force balance gives:
\begin{equation}
    \sum F = T\sin(\theta)-F_D = ma,
\label{eq1}
\end{equation}
where $T$ is the filament force, $\theta$ is the angle between the pulling direction and the bead displacement, $m$ is the bead mass, and $a$ is its acceleration. The viscous drag force is given by Stokes' law:
\begin{equation}
    F_D=-6\pi\eta r v,
\label{eq2}
\end{equation}
where $\eta$ is the medium viscosity, $r$ is the bead radius, and $v$ is the bead velocity.

For rotational motion, the filament force generates a torque that balances the viscous rotational drag:
\begin{equation}
    Tr\cos(\theta)=\tau_D=8\pi\eta r^3\omega,
\label{eq3}
\end{equation}
where $r$ is the distance from the bead center to the point of force application, $\omega$ is the angular velocity, $\theta$ is the angle of the filament to the bead, and $\tau_D$ is the viscous drag torque. Analysis of both translational and rotational events revealed that individual filament arms exert a force of $68.3 \pm 46.7$ pN (mean $\pm$ SD, $n=7$ events, Table S1).

The force measurements presented here establish a direct link between single-filament mechanics and emergent collective behavior. We propose a hierarchical mechanism for force generation within the cyanobacterial network: at the microscopic scale, individual filament contact points retract, generating forces in the pN range. At the single-contact scale, chiral wrapping, the right-handed bending of each filament around its partner, may amplify this force through geometric effects. Finally, at the network scale, multiple simultaneous contacts sum to produce cumulative contractile forces. A simple scaling estimate suggests that with $n_c \sim 10^3\text{ contacts mm}^{-2}$, each contributing $\sim 100\text{ pN}$, the network generates a cumulative stress of $\sigma \sim 100\text{ nN mm}^{-2}$, which is sufficient to overcome viscous drag and drive contraction on the timescale of minutes.

\subsection*{Network Model of Contraction Dynamics}

To test whether the measured single-contact force ($68.3 \pm 46.7$ pN, approximately assumed as 100 pN) can account for the observed macroscopic contraction dynamics, we developed a coarse-grained network model that bridges molecular-scale forces to network-level behavior. The model serves as a quantitative test of the hypothesis that filament-filament contacts, when integrated across the network, are sufficient to drive the rapid contraction observed experimentally.

\begin{figure}[t!]
    \centering
    \includegraphics[width=\linewidth]{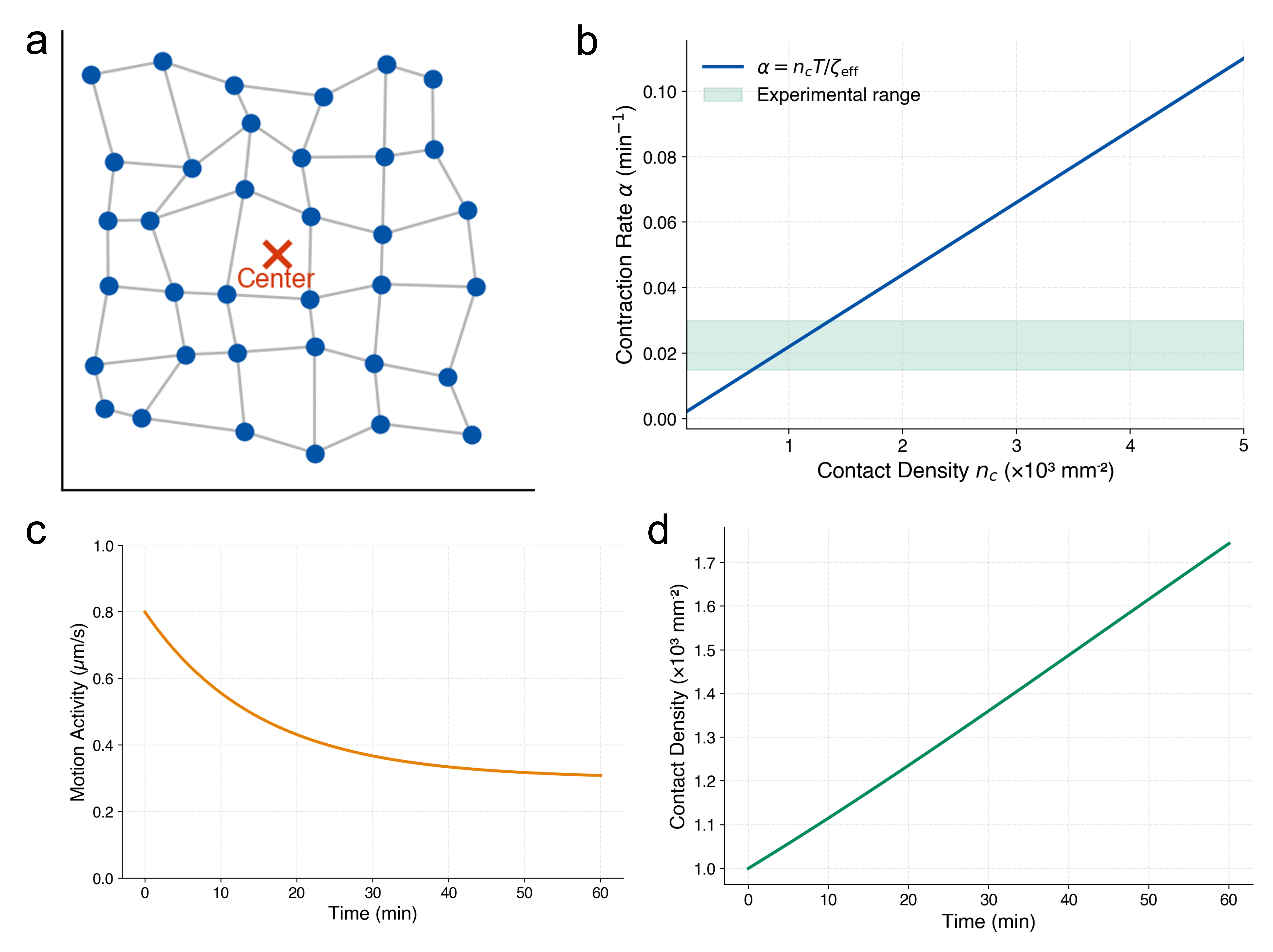}
    \caption{Network model linking single-contact forces to macroscopic contraction. a) Schematic of the coarse-grained filament network with nodes (circles) connected by springs (gray lines), and inward contractile stress toward the center (red cross). b) Predicted contraction rate $\alpha$ as a function of contact density $n_c$ from $\alpha = n_cT/\zeta_{\mathrm{eff}}$ using $T = 100$ pN. The red point marks $n_c \approx 10^3\text{ mm}^{-2}$, with the shaded region indicating the experimental range. c) Simulated motion activity decreases over time as elastic forces balance contractile stress. d) Simulated contact density increases from $\sim 1 \times 10^3$ to $1.76 \times 10^3\text{ mm}^{-2}$ during contraction.}
    \label{fig55}
\end{figure}

The filament network is represented as a set of $N$ nodes (Figure~\ref{fig55}a, filament intersections) connected by elastic springs (filament segments). Each node $i$ at position $\mathbf{r}_i$ experiences three types of forces. First, elastic restoring forces arise from filament segments modeled as Hookean springs with spring constant $k$, current distance $d_{ij}$, and rest length $l_{ij}^0$~\cite{cordoba2014single}. The elastic force on node $i$ due to its connection to node $j$ is:
\begin{equation}
F_{ij}^e = k \left( d_{ij} - l_{ij}^0 \right) \frac{\mathbf{r}_j - \mathbf{r}_i}{d_{ij}}
\label{eq:elastic}
\end{equation}
The spring constant $k \sim 10^{-6}\text{ N m}^{-1}$ is estimated from the bending stiffness of cyanobacterial filaments~\cite{faluweki2022structural}. 

Second, contractile forces arise from filament-filament contacts. At the coarse-grained level, we treat the network as an isotropic contractile medium. For a network with contact density $n_c$ (number of contacts per unit area), each contact transmitting force $T$, the total contractile stress is $\sigma_c = n_c T$. The contractile force on node $i$ is:
\begin{equation}
    \mathbf{F}_i^{\text{con}} = \alpha_c \, n_c T \, (\mathbf{r}_c - \mathbf{r}_i) \, A_i
    \label{eq:contractile}
\end{equation}
where $\mathbf{r}_c$ is the network centroid, $A_i$ is the Voronoi area associated with node $i$, and $\alpha_c$ is a geometric factor. This formulation represents a mean-field approximation: although individual filament-filament contacts are local and randomly oriented, their collective effect over the network generates a net inward stress field. The contact density is estimated from the network geometry as $n_c \sim 10^3\text{ mm}^{-2}$, corresponding to a mesh-like network with bundle spacing of approximately 30~$\mu\text{m}$.

Third, motion through the network is opposed by effective friction. Unlike a single filament moving through a Newtonian fluid, filaments within a dense network experience resistance primarily through filament-filament sliding, transient adhesions, steric interactions, entanglement, and topological constraints associated with helical tangles. We therefore introduce an effective friction coefficient $\zeta_{\text{eff}}$ that captures all these dissipation mechanisms. The friction force on a node is given by:
\begin{equation}
    \mathbf{F}_i^{\text{friction}} = -\zeta_{\text{eff}} \frac{d\mathbf{r}_i}{dt}
    \label{eq:friction}
\end{equation}

The dynamics of each node are governed by overdamped motion:
\begin{equation}
    \zeta_{\text{eff}} \frac{d\mathbf{r}_i}{dt} = \sum_{j} \mathbf{F}_{ij}^{\text{e}} + \mathbf{F}_i^{\text{con}}
    \label{eq:motion}
\end{equation}

The model admits a simple scaling analysis that provides physical intuition. Because macroscopic contraction is driven by active forces and resisted by internal dissipation, balancing the active contractile stress against network frictional resistance gives:
\begin{equation}
    \sigma_c \sim \zeta_{\text{eff}} \dot{\varepsilon}
    \label{eq:scaling_balance}
\end{equation}
where $\dot{\varepsilon}$ is the contraction strain rate representing the rate of network deformation. Substituting the active stress relation $\sigma_c = n_c T$ (where $n_c$ is the contact density and $T$ is the single-contact force) and expressing the strain rate in terms of the macroscopic contraction rate $\alpha \equiv -\dot{R}/R$ (where $R$ is the network radius), we obtain:
\begin{equation}
    \alpha \sim \frac{n_c T}{\zeta_{\text{eff}}}
    \label{eq:scaling_alpha}
\end{equation}

This expression dimensionally captures the essential physics: contraction results from a balance between active stress generation and network friction. Using our measured values ($n_c = 10^3\text{ mm}^{-2}$, $T = 1.0 \times 10^{-10}$ N, and $\alpha \sim 0.022\text{ min}^{-1}$ from experiments, Appendix Figure 6), we obtain $\zeta_{\text{eff}} \sim 2.7 \times 10^{-1}\text{ Pa}\cdot\text{s}$ (Figure~\ref{fig55}b). This value is approximately 270 times larger than water viscosity ($10^{-3}\text{ Pa}\cdot\text{s}$), which is physically reasonable for a dense network where filament-filament interactions and topological constraints provide substantial additional resistance.

Using the estimated effective friction coefficient, the predicted contraction rate is $\alpha \sim \frac{n_c T}{\zeta_{\text{eff}}} \times 60\text{ s min}^{-1} \sim 2.2\text{ min}^{-1}$, which is approximately two orders of magnitude higher than the experimentally observed rate of $\sim 0.022\text{ min}^{-1}$. This overestimation is expected given the simplicity of the mean-field model, which assumes all contacts contribute uniformly while neglecting heterogeneous orientations and intermittent interactions. Nevertheless, the scaling analysis serves as a proof of principle that pN-scale forces are sufficient to drive macroscopic contraction.

To further test the consistency of the model, we numerically integrated Eq.~\ref{eq:motion} using the estimated parameters and monitored both the average nodal velocity and the evolving contact density during contraction (Figure~\ref{fig55}c,d). The mean motion activity, defined as the average node speed, decreases rapidly from an initial value of $\sim 0.8~\mu\text{m}\cdot\text{s}^{-1}$ to a plateau of $\sim 0.3~\mu\text{m}\cdot\text{s}^{-1}$ over the course of the simulation (Figure~\ref{fig55}c). At the same time, the contact density increases monotonically from approximately $1\times10^{3}$ to $1.76\times10^{3}\text{ mm}^{-2}$ (Figure~\ref{fig55}d), consistent with the geometric expectation that contraction brings filaments into closer proximity. The increase in $n_c$ provides a positive feedback on contractility through Eq.~\ref{eq:scaling_alpha}, while being progressively balanced by elastic resistance and frictional dissipation, preventing runaway collapse and leading to a steady contraction rate comparable to experiments.

\section*{Conclusions}

In this study, we uncovered a physical mechanism by which chiral gliding drives the self-organization of filamentous cyanobacteria, \textit{Lyngbya lagerheimii}, into compact, densely packed aggregates. By examining these active filaments across multi-scale environments, we observed that individual filaments do not merely glide passively alongside one another; rather, they actively intertwine through distinct helical tangling interactions. This entanglement process generates measurable contractile forces at single-contact points. Remarkably, these pN-scale forces, when integrated across the network, produce sufficient cumulative stress ($\sigma \sim 100\text{ nN mm}^{-2}$) to drive rapid macroscopic contraction.

Our findings establish a direct, quantitative link between single-filament chirality and emergent collective behavior in active matter. The chiral rotation of each filament promotes sustained entanglement, stabilizes robust bundles, and enables the efficient transmission of contractile tension across the network. We propose that this entanglement-driven mechanism represents an effective strategy that facilitates self-organization, allowing cyanobacteria to form cohesive mats in aquatic environments despite environmental perturbations. From a broader perspective, our work provides a physical framework for understanding how active, chiral, and flexible filamentous systems achieve large-scale organization through local interactions. Furthermore, the entanglement-mediated contraction mechanism uncovered here may inspire the design of synthetic active materials that harness chiral interactions for controlled assembly, actuation, and mechanical adaptation.

\section*{Methods and Materials}

\subsection*{Bacteria culture}
\textit{Lyngbya lagerheimii} was obtained from The Culture Collection of Algae at Göttingen University (SAG strain number 24.99) and maintained in T175 culture flasks with standard BG-11 medium (Gibco™, ThermoFisher). The culture medium was exchanged with fresh BG-11 every four weeks. Cultures were kept in an incubator with an automated 12 h day (30\% light intensity, 20 $\mu E$, 18$^{\circ}$C) and 12 h night (0\% light intensity, 14$^{\circ}$C) cycle, with a continuous 2 h transition period between cycles.

\subsection*{Quasi-two-dimensional confinement}
Cyanobacterial filaments were confined between two high-resolution microscope glass slides. A 100 $\mu\text{m}$ thick double-sided adhesive tape was used as a spacer to create a precisely defined observation chamber, effectively restricting filament movement to a quasi-two-dimensional plane. For larger-scale observations, a custom-built chamber with dimensions of $1.5\text{ cm} \times 1.5\text{ cm} \times 0.1\text{ cm}$ was used.

\subsection*{Microscopy and image acquisition}
Microscopic observations were conducted using two imaging systems. For 2D confinement observations at high magnification, an Olympus CKX41 inverted microscope equipped with an Axiocam MRm camera and phase contrast optics was used. For wider-field visualization of contraction-expansion cycles, a Zeiss Axio Zoom V.16 microscope equipped with a Hamamatsu ORCA-Flash 4.0 digital camera and a Zeiss PlanNeoFluar 1×/0.25 objective was employed. Time-lapse images were acquired at 1 min intervals for up to 12 h.

\subsection*{Image analysis and texture quantification}
Image analysis was performed using custom Python scripts. The pixel-to-micron calibration was determined from the imaging setup. For each frame, images were converted to grayscale and normalized to 8-bit format.

Filament alignment coherence was quantified using the structure tensor method. For each image, the local gradient orientation and magnitude were computed using Sobel filters with a kernel size of 3. The structure tensor $J$ was calculated as:
\begin{equation}
J = \begin{pmatrix}
\langle I_x^2 \rangle & \langle I_x I_y \rangle \\
\langle I_x I_y \rangle & \langle I_y^2 \rangle
\end{pmatrix}
\end{equation}
where $I_x$ and $I_y$ are the image gradients in the $x$ and $y$ directions, and $\langle \cdot \rangle$ denotes averaging over a local window of size $15 \times 15$ pixels using a uniform kernel. The eigenvalues of the structure tensor, $\lambda_1$ and $\lambda_2$, were computed as:
\begin{equation}
\lambda_{1,2} = \frac{J_{xx} + J_{yy} \pm \sqrt{(J_{xx} - J_{yy})^2 + 4J_{xy}^2}}{2}
\end{equation}
The coherence $C$ was then defined as:
\begin{equation}
C = \frac{\lambda_1 - \lambda_2}{\lambda_1 + \lambda_2 + \epsilon}
\end{equation}
where $\epsilon = 10^{-6}$ prevents division by zero. Coherence values range from 0 (isotropic texture, randomly oriented filaments) to 1 (perfectly aligned filaments). The straightening rate was computed as the temporal derivative of coherence: $dC/dt$, representing the rate of change in filament alignment per minute.

Motion activity was quantified using optical flow analysis with the Farnebäck algorithm~\cite{farneback2003two}, implemented in OpenCV. The mean motion speed was calculated as the average magnitude of the optical flow field, converted to $\mu\text{m}\cdot\text{s}^{-1}$ using the pixel-to-micron calibration and the 1 min time interval between frames:
\begin{equation}
v = \frac{\langle |\mathbf{v}_{\text{flow}}| \rangle \times (1 / 0.18)}{60 \text{ s}}
\end{equation}

Motion coherence was computed using circular statistics on the flow directions, weighted by magnitude. The curvature-motion correlation, defined as the correlation between vorticity and divergence of the flow field, was used as a proxy for rotational (tangling) versus translational (sliding) motion.

\subsection*{Fabrication of cyanobacteria-encapsulated alginate beads}
Cyanobacterial filaments were encapsulated within alginate beads through a stepwise process. Sodium alginate solutions were prepared at concentrations of 1\% and 2\% (w/v) in BG-11 medium. Cyanobacterial culture was mixed with alginate solution in a 1:1 ratio (1 mL total volume) and subjected to vigorous vortexing for 30 s to ensure homogeneous mixing. The resulting cyanobacterial-alginate mixture (5 $\mu\text{L}$ droplets) was then carefully dispensed using a micropipette onto a 3\% (w/v) calcium chloride solution, triggering gelation and forming spherical cyanobacteria-encapsulated alginate beads (CABs). Beads were transferred to fresh BG-11 medium and maintained at room temperature under ambient light conditions.

\subsection*{Force measurement from bead displacement}
The force exerted by a single filament arm was determined by tracking CAB displacement from time-lapse images acquired at 30 s intervals. Bead trajectories were analyzed using ImageJ (NIH) with the Manual Tracking plugin. For each bead, the position as a function of time was extracted, and velocity and acceleration were computed by numerical differentiation.

\subsection*{Network model simulation}
The coarse-grained network model was implemented in Python (NumPy, SciPy). The filament network was represented as a set of $N = 64$ nodes arranged on a hexagonal lattice, connected by springs with spring constant $k = 10^{-6}\text{ N m}^{-1}$. Nodes were initialized with random perturbations of 5\% of the lattice spacing to break symmetry. The equations of motion were integrated using a forward Euler scheme with time step $\Delta t = 0.1$ s for a total simulation time of 600 s. Contractile forces were applied according to Eq.~\ref{eq:contractile} with $T = 1.33 \times 10^{-12}$ N, $n_c = 10^3\text{ mm}^{-2}$, and $\alpha_c = 0.01$. The effective friction coefficient was set to $\zeta_{\text{eff}} = 2.7 \times 10^{-3}\text{ Pa}\cdot\text{s}$ based on the scaling estimate. Contact density was computed at each time step as the number of node pairs within a distance of 1.2 times the initial lattice spacing.

\subsection*{Statistical analysis}
All data are reported as mean $\pm$ standard deviation (SD) unless otherwise noted. 

\section*{Data availability}
All analysis codes, numerical scripts, and the network model implementation used in this study have been deposited in Figshare under \url{https://doi.org/10.6084/m9.figshare.33173960}. Source data corresponding to the measurements and simulations presented in the figures are available within the repository.

\section*{Acknowledgments}
We thank Diana Strüver for conducting the bacteria culture. The authors thank Robab Jahangir for helping with the alginate bead preparation and conducting the bead experiments.

\section*{Additional information}

\subsection*{Author contributions}
V.N. designed research; V.N and J.Y. conducted the experiment; V.N., J.Y and H.K analyzed data; V.N. developed the theoretical model and performed simulations; V.N., J.Y. and H.K wrote the paper.

\subsection*{Funding}
This work was supported by the Max Planck Society and the Max Planck Partner Group Program of the Max Planck Society (Q2322361).

\bibliographystyle{unsrt}
\bibliography{bib}

\clearpage

\appendix
\section*{Supplementary Information}
\setcounter{figure}{0}
\renewcommand{\thefigure}{S\arabic{figure}}

\setcounter{table}{0}
\renewcommand{\thetable}{S\arabic{table}}

\subsection*{Experimental Setup}
To confine cyanobacterial filaments to a quasi-two-dimensional environment, we constructed a microfluidic chamber using a double-sided adhesive tape spacer. The chamber consisted of two glass microscope slides separated by a 100 $\mu\text{m}$ thick tape layer that served as both a spacer and bonding material. The tape was pre-cut to create a rectangular observation window ($\sim 1.5\text{ cm} \times 1.5\text{ cm}$) defining the active area for filament confinement. A 10 $\mu\text{L}$ droplet of cyanobacterial culture in BG-11 medium was introduced into the chamber.

The 100 $\mu\text{m}$ chamber height was larger than the filament diameter ($\approx 4~\mu\text{m}$) but small enough to restrict out-of-plane motion, creating a quasi-2D environment. This design enabled continuous time-lapse imaging over extended periods (up to 24 h) while maintaining filaments in a single focal plane for quantitative analysis of alignment, motion, and contraction dynamics.

\begin{figure}[htbp]
\centering
\includegraphics[width=0.6\linewidth]{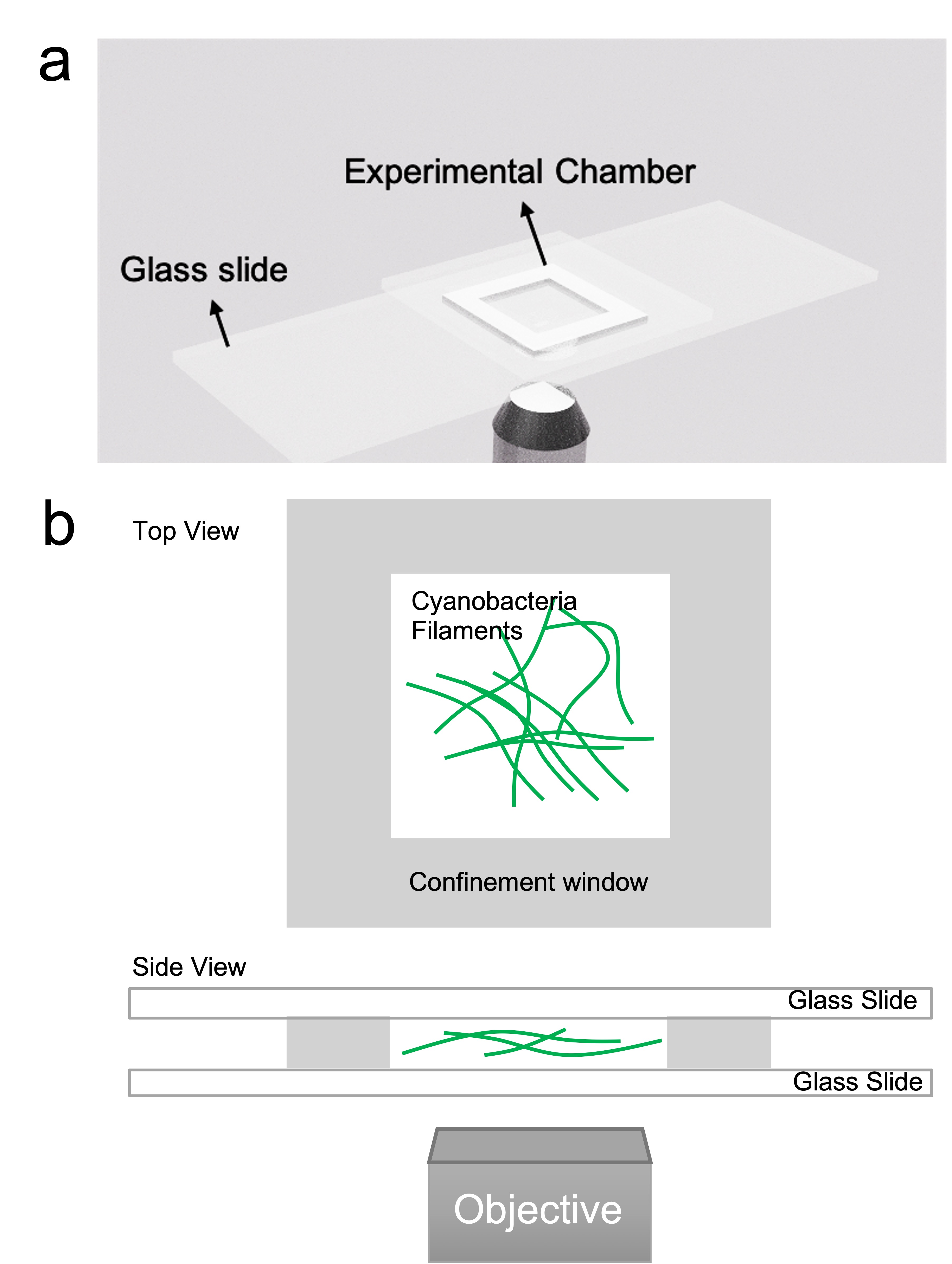}
\caption{Experimental chamber for quasi-two-dimensional confinement. a) 3D schematic showing the glass slides separated by a 100 $\mu\text{m}$ thick adhesive tape spacer with the observation window. b) Top and side views of the chamber, illustrating the observation area and the 100 $\mu\text{m}$ height that restricts filaments to a single focal plane.}
\label{fig:S3}
\end{figure}

\subsection*{Oxygen Bubble Nucleation and Growth}
In the confined quasi-two-dimensional chamber, the photosynthetic activity of the cyanobacterial filaments led to the nucleation and growth of oxygen bubbles. The confinement geometry constrained bubble expansion in the vertical direction, forcing them to grow laterally within the imaging plane. Once nucleated, the bubbles expanded rapidly over tens of minutes, displacing surrounding filaments and locally disrupting the filamentous network. The growth dynamics followed a characteristic pattern: initial bubble nucleation at preferential sites such as filament intersections or impurities, followed by gradual expansion as oxygen production continued. As the bubbles grew, they acted as physical obstacles, impeding network contraction and creating localized regions of filament depletion. In some cases, bubble growth was sufficient to detach filaments from the substrate or induce local rearrangements of the network.

\begin{figure}[htbp]
\centering
\includegraphics[width=\linewidth]{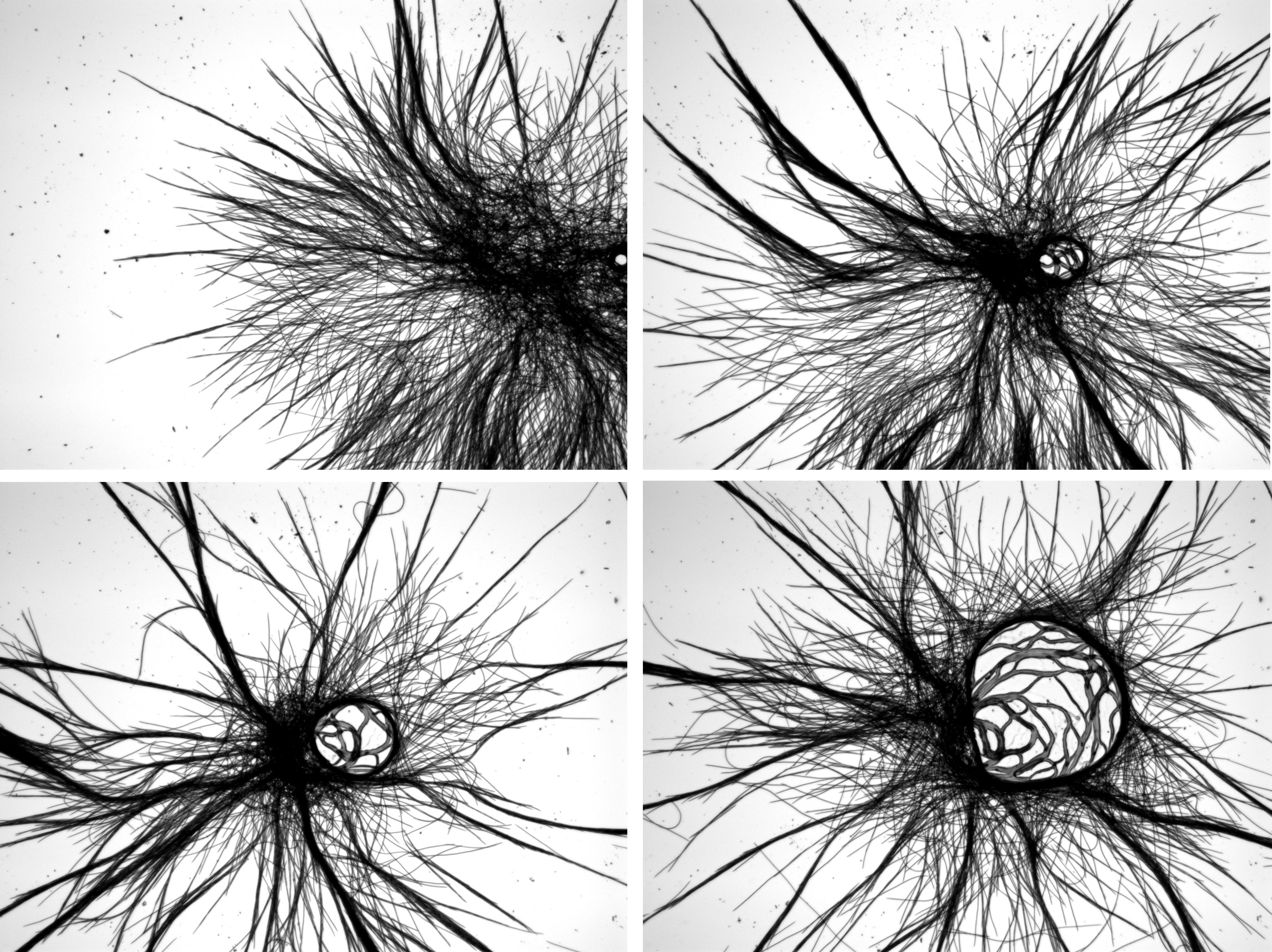}
\caption{Time-lapse images showing the nucleation and growth of an oxygen bubble within the quasi-2D confinement. The bubble nucleates at a filament intersection ($t = 0$ min) and expands over time ($t = 15, 30, 45$ min), displacing surrounding filaments and creating a localized region of network disruption. The confinement restricts bubble growth to the imaging plane, resulting in lateral expansion.}
\label{fig:S2}
\end{figure}

\subsection*{Chiral Gliding of a Short Filament Around a Longer Filament}
To further investigate the role of chiral gliding in filament-filament interactions, we examined a configuration in which a longer filament was clamped to the glass substrate at one end while its other end remained free, and a much shorter filament landed on the longer filament. Time-lapse microscopy revealed that once the short filament made contact with the longer filament, it did not simply glide alongside it. Instead, driven by its intrinsic chiral gliding motion, the short filament began to rotate around the longer filament, tracing a helical path along its length. This rotational motion is a direct consequence of the filament's clockwise rotation about its long axis during gliding, which causes it to follow a curved trajectory when in contact with another filament. The longer filament served as a stationary substrate, while the short filament's chiral motion caused it to wind around the longer filament in a helical fashion. Over time, the short filament completed multiple turns around the longer filament, resulting in a stable helical tangle. This observation confirms that chiral gliding is a sufficient mechanism for generating helical tangles between filaments, even without active gliding of both partners. The experiment also demonstrates that the rotation is not a consequence of mutual entanglement but rather an intrinsic property of the chiral gliding motion itself.

\begin{figure}[htbp]
\centering
\includegraphics[width=\linewidth]{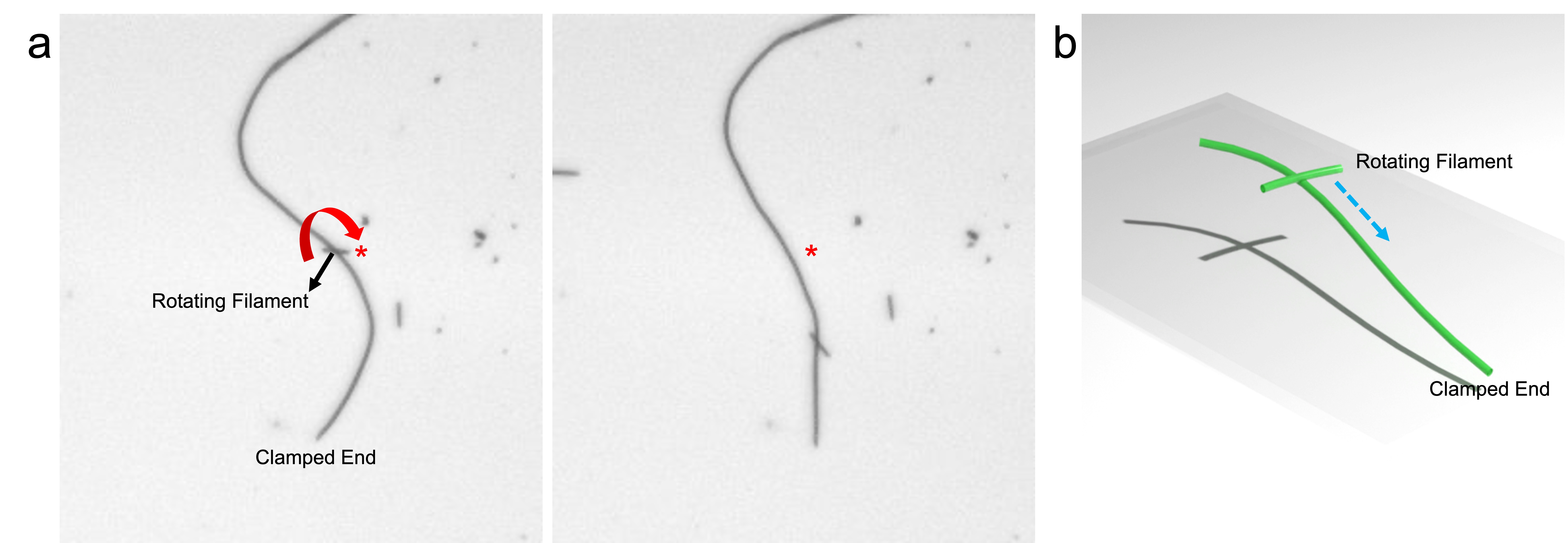}
\caption{Chiral gliding of a short filament around a longer, clamped filament. a) Time-lapse microscopy showing a shorter filament (arrowhead) landing on a longer filament (arrow) that is clamped to the substrate at one end. The short filament rotates around the longer filament, tracing a helical path along its length. b) Schematic illustration of the experimental configuration: the longer filament is fixed at one end while the short filament glides and rotates around it due to its intrinsic chiral motion. This demonstrates that chiral gliding alone is sufficient to generate helical tangles without requiring mutual active motion of both partners.}
\label{fig:S5}
\end{figure}

\subsection*{Filament Bundling Through Tangling Interactions}
The formation of filament bundles is a key step in the self-organization of the cyanobacterial network, and tangling interactions play a central role in bringing individual filaments together into thicker, mechanically stable bundles. Time-lapse microscopy reveals a characteristic sequence of events leading to bundle formation. Multiple filaments often making transient contact at multiple points along their lengths. These contacts are stabilized by the adhesive properties of the extracellular polymeric substance (EPS) sheath surrounding each filament. As the filaments continue to glide, the contacts increase in number and the filaments begin to wrap around one another in a helical fashion, similar to the helical tangles observed in the contracting network (Figure~\ref{fig3}). The formation of bundles reduces the effective number of independent filaments in the network, concentrating the contractile forces generated by individual contacts and thereby enhancing the overall mechanical cohesion of the network.

\begin{figure}[t!]
\centering
\includegraphics[width=\linewidth]{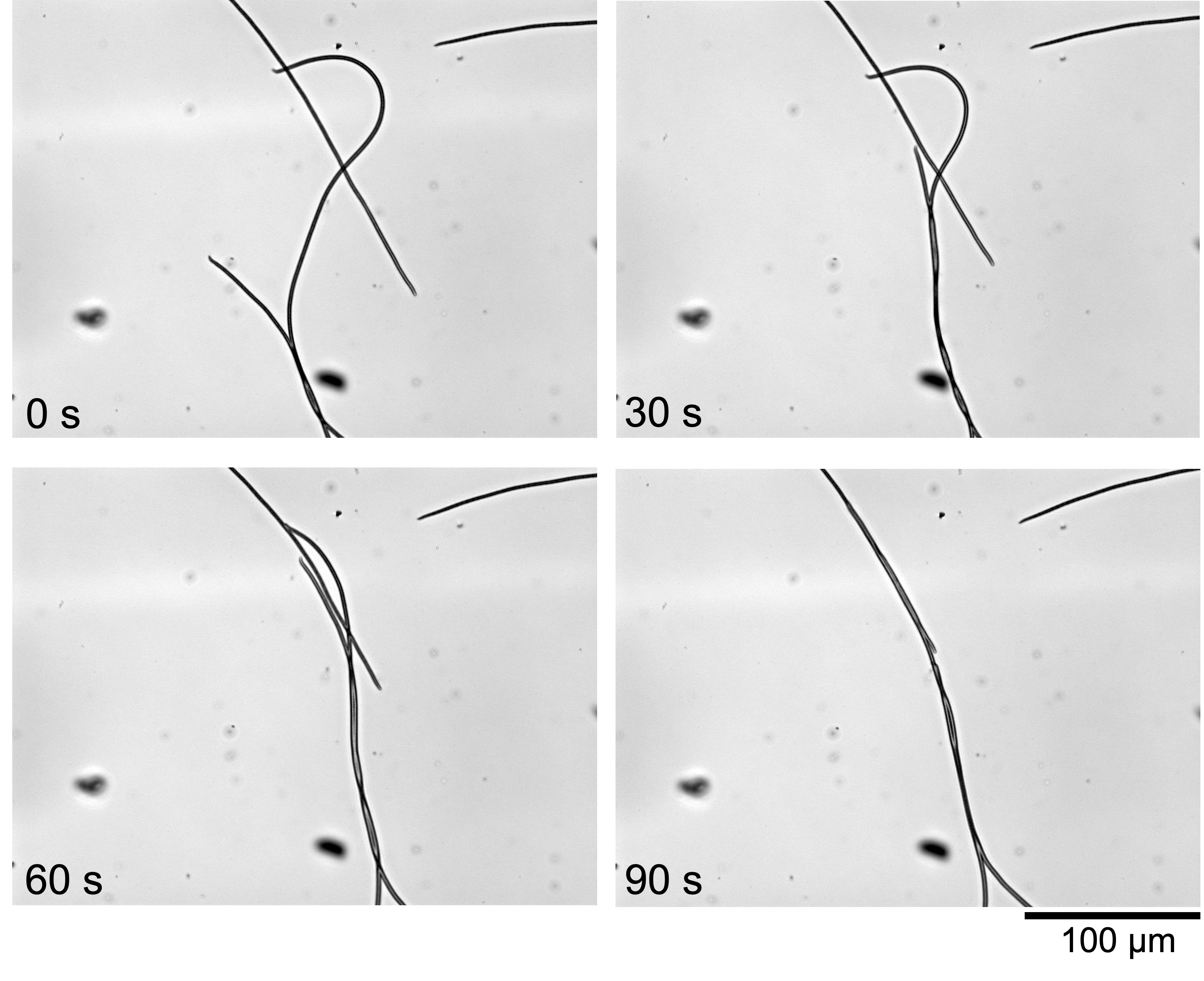}
\caption{Time-lapse microscopy showing filament bundling through tangling interactions. Three individual filaments (arrows) glide in close proximity and engage in multiple filament-filament contacts ($t = 0$ min). Over time, the filaments wrap around one another in a helical fashion ($t = 5, 10$ min), eventually forming a single compact bundle ($t = 15$ min).}
\label{fig:S1}
\end{figure}

\subsection*{Force Analysis Method}
The force exerted by a single cyanobacterial filament arm on an alginate bead was quantified using time-lapse imaging and force balance analysis. Bead trajectories were tracked manually using ImageJ, and the position, velocity, and acceleration of each bead were extracted from the displacement data. The mass of each cyanobacteria-encapsulated alginate bead (CAB) was calculated from its volume and the densities of its constituent components (sodium alginate, BG-11 medium, and cyanobacterial biomass). The density of BG-11 medium was estimated from the weighted contributions of its components, assuming the salt volume was negligible compared to water:
\begin{equation}
    \rho_{\text{BG-11}} = \frac{\sum m_i}{V_{\text{water}}}
\end{equation}
where $\sum m_i$ is the total mass of all BG-11 components and $V_{\text{water}}$ is the volume of water. The cyanobacterial culture solution was assumed to consist of 50\% BG-11 medium and 50\% cyanobacterial biomass (density $\rho_{\text{biomass}} \approx 1.05\text{ g cm}^{-3}$).

Depending on the interaction geometry, the filament induced either translational or rotational motion of the bead. For translational (pulling) motion, the force balance is given by:
\begin{equation}
    \sum F = T \sin(\theta) + F_D = m a
\end{equation}
where $T$ is the tension force exerted by the filament arm, $\theta$ is the angle between the direction of bead movement and the applied force, $m$ is the bead mass, and $a$ is the bead acceleration. The drag force $F_D$ was calculated using Stokes' law:
\begin{equation}
    F_D = -6\pi \eta r v
\end{equation}
where $\eta = 10^{-3}\text{ Pa}\cdot\text{s}$ is the medium viscosity (assumed same as BG-11), $r$ is the bead radius, and $v$ is the bead velocity. 

For the bead rotation case, under the low Reynolds number regime ($\text{Re} \sim 10^{-6}$), fluid inertia is negligible and the motion is quasi-steady, meaning bead angular inertia can be neglected ($I \frac{d\omega}{dt} = 0$). The torque balance accounts for the filament tension torque balanced by the Stokes' rotational drag torque:
\begin{equation}
    \sum \text{Torque} = T \cdot \cos(\theta) \cdot r - \tau_D = 0
\end{equation}
where the rotational drag torque is given by:
\begin{equation}
    \tau_D = 8\pi\eta r^3 \omega
\end{equation}
with $\omega$ representing the angular velocity, $r$ the bead radius, and $\theta$ the angle between the applied force and the bead movement direction. Rearranging this torque balance yields the tension calculation for rotation:
\begin{equation}
    T = \frac{8\pi\eta r^3 \omega}{\cos(\theta) \cdot r}
\end{equation}

The calculated material properties and force measurements for both rotational and pulling cases are summarized in Tables~\ref{tab:force_measurements} and~\ref{tab:material_properties}.

\begin{table}[htbp]
\centering
\caption{Measured drag force ($F_D$), drag torque ($\tau_D$), and tension ($T$) for CEB samples under rotation and pulling cases.}
\label{tab:force_measurements}
\vspace{0.5em}
\begin{tabular}{llcc}
\toprule
Motion Type & Sample & Drag ($\tau_D$ [pN$\cdot\mu\text{m}$] or $F_D$ [pN]) & Tension $T$ (pN) \\
\midrule
\textbf{Bead rotation} 
& CEB\_2\% & $0.066$ ($\tau_D$) & $161.78$ \\
& CEB\_2\% & $0.044$ ($\tau_D$) & $46.81$ \\
& CEB\_2\% & $0.057$ ($\tau_D$) & $76.09$ \\
& CEB\_1\% & $0.047$ ($\tau_D$) & $78.90$ \\
& CEB\_1\% & $0.038$ ($\tau_D$) & $63.89$ \\
\midrule
\textbf{Bead pulling} 
& CEB\_2\% & $16.20$ ($F_D$) & $16.31$ \\
& CEB\_1\% & $35.61$ ($F_D$) & $35.79$ \\
\bottomrule
\end{tabular}
\end{table}

\begin{table}[htbp]
\centering
\caption{Density, volume, and mass of materials used in the experiments.}
\label{tab:material_properties}
\vspace{0.5em}
\begin{tabular}{lccc}
\toprule
Material & Density ($\text{g cm}^{-3}$) & Volume (mL) & Mass (g) \\
\midrule
BG-11 & 1.001687723 & -- & -- \\
1 wt.\% alginate solution & 1.011687723 & 0.5 & 0.505844 \\
2 wt.\% alginate solution & 1.021687723 & 0.5 & 0.510844 \\
Cyanobacteria & 0.87 & -- & -- \\
Cyanobacteria solution & -- & 0.5 & 0.467922 \\
CEB\_1\% & 0.973766 & 0.005 & 0.004869 \\
CEB\_2\% & 0.978766 & 0.005 & 0.004894 \\
\bottomrule
\end{tabular}
\end{table}

\subsection*{Bead Deformation as the Result of Filament-Filament Interactions}

\begin{figure}[htbp]
\centering
\includegraphics[width=\linewidth]{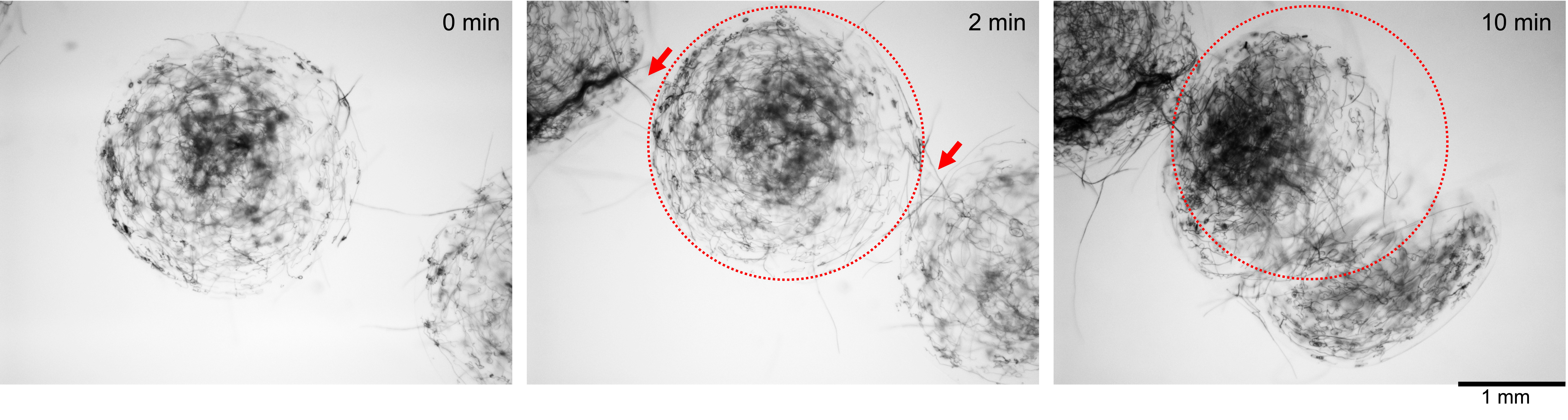}
\caption{The time-lapse images show connecting of the beads through the extension arms and eventual deformation of the beads due to applied forces by the filament.}
\label{fig:S6}
\end{figure}

\newpage
\subsection*{Contraction Node Analysis and Contraction Rate Derivation}
To link microscopic node dynamics to the macroscopic contraction rate $\alpha$, we use the geometric relationship between node density and the network radius. The contraction rate is defined as:

\begin{equation}
    \alpha = -\frac{1}{R} \frac{dR}{dt}
\end{equation}
where $R$ is the network radius. For uniform two-dimensional contraction, the node density $n$ is inversely proportional to the network area, which scales with the square of the radius ($n \propto \text{Area}^{-1} \propto R^{-2}$). Taking the natural logarithm and differentiating with respect to time yields:
\begin{equation}
    \frac{1}{n} \frac{dn}{dt} = -2 \left(\frac{1}{R} \frac{dR}{dt}\right) = 2\alpha
\end{equation}
Thus, the macroscopic contraction rate is determined directly from the relative node increase rate evaluated at the midpoint of the temporal data:
\begin{equation}
    \alpha = \frac{1}{2} \left( \frac{\text{slope}}{N} \right)
\end{equation}
where $\text{slope} = dn/dt$ from the linear fit of node count over time, and $N$ is the node count evaluated at the temporal midpoint.

\begin{figure}[t!]
\centering
\includegraphics[width=\linewidth]{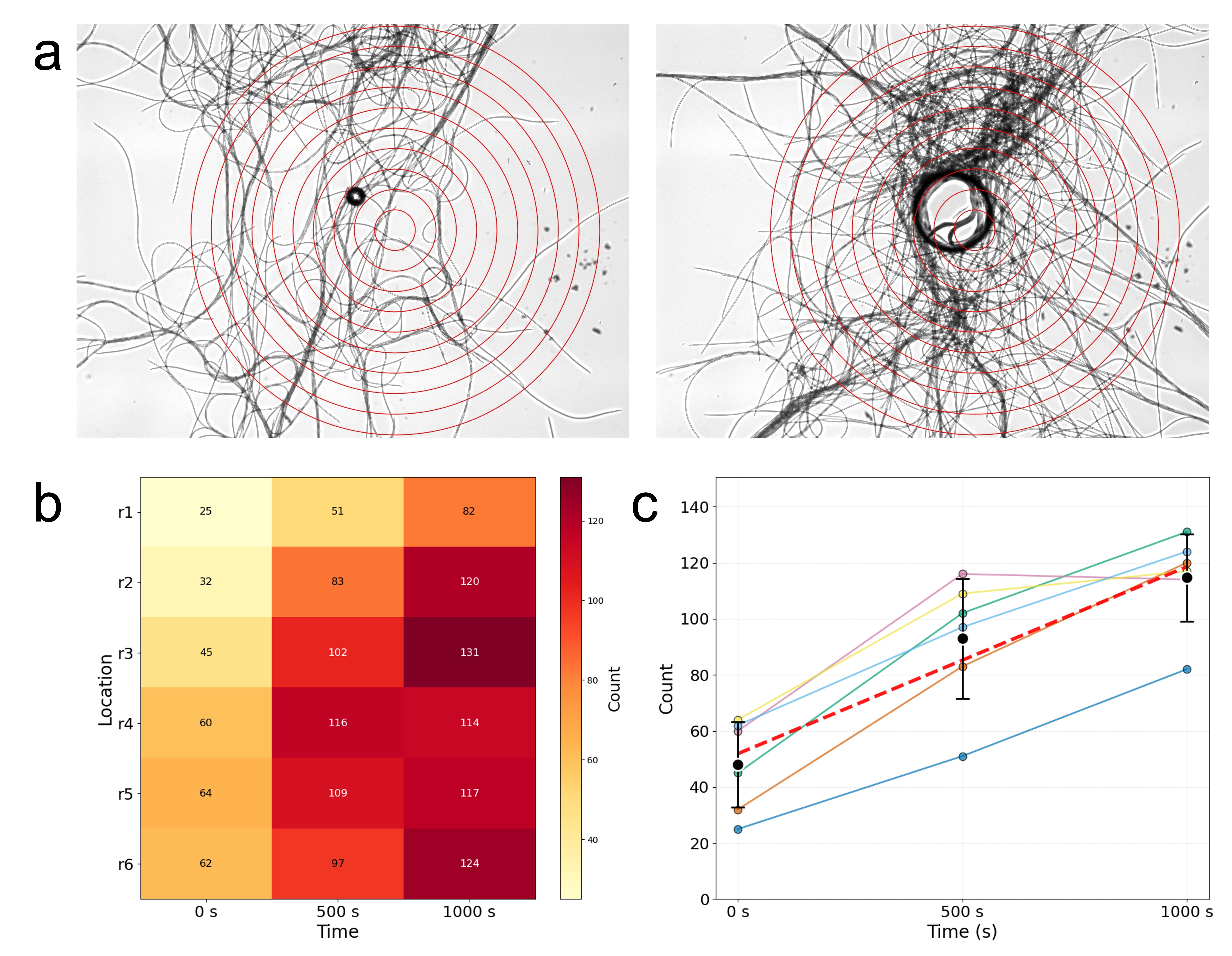}
\caption{Contraction behavior of the cyanobacterial filament network over time. a) Microscopic images showing network evolution with inner (light blue) and outer (yellow) regions highlighted. b) Total number of nodes in the region of interest over time. c) Number of nodes inside the inner region. d) Number of nodes at the outer region, used for extracting linear growth rates to calculate the contraction parameter $\alpha$.}
\label{fig:S4}
\end{figure}

\end{document}